\documentclass[aps,twocolumn,secnumarabic,
               nobalancelastpage,amsmath,amssymb,pre]{revtex4}
\usepackage{mathrsfs}
\usepackage{physics}
\usepackage{hyperref}
\usepackage{bm}
\usepackage{url} 
\usepackage{graphicx}
\usepackage{subfigure}
\usepackage{xcolor}

\newcommand{\ba}{\begin{equation}\begin{aligned}}
\newcommand{\ea}{\end{aligned}\end{equation}}

\newcommand{\matr}[1]{\bm{#1}}

\usepackage{bm}
\usepackage{bbm}
\newcommand{\W}{\mathcal{W}}

\begin{document}
\title{Dynamical phase behavior of the  single- and multi-lane asymmetric 
       simple exclusion process via matrix product states}
\author{Phillip Helms}
\email{phelms@caltech.edu}
\author{Ushnish Ray}
\email{uray@caltech.edu}
\author{Garnet Kin-Lic Chan}
\email{garnetc@caltech.edu}
\affiliation{Division of Chemistry and Chemical Engineering, 
             California Institute of Technology, Pasadena,
             CA 91125}
\date{\today}
\begin{abstract}
  We analyze the dynamical phases of the current-biased 1D and multi-lane open 
    asymmetric simple exclusion processes (ASEP),
  using matrix product states and the density matrix renormalization group 
    (DMRG) algorithm. 
  In the 1D ASEP, we present a systematic numerical study of the current 
    cumulant generating function and its derivatives, 
  which serve as dynamical phase order parameters.
  We further characterize the microscopic structure of the phases from 
    local observables and the entanglement spectrum. 
  In the multi-lane ASEP, which may be viewed as finite width 2D strip, 
    we use the same approach and 
  find the longitudinal current-biased dynamical phase behavior to be 
    sensitive to transverse boundary conditions. 
  Our results serve to illustrate the potential of tensor networks in 
    the simulation of classical nonequilibrium statistical models.
\end{abstract}
\maketitle

\section{Introduction}

Connecting microscopic states to macroscopic properties is a central goal of 
statistical mechanics. At equilibrium, this connection is 
expressed through the Gibbs-Boltzmann framework, which defines the free energy 
and its derivatives in terms of microscopic states.
Large deviation theory (LDT) provides an analogous framework for nonequilibrium
systems. Large deviation functions (LDFs), 
such as the cumulant generating function (CGF) $\psi$, and the rate function 
$\phi$,  are analogs of the equilibrium free energy and entropy. 
Their derivatives provide information on dynamical order parameters and rare 
fluctuations as a system is driven away from equilibrium
~\cite{Donsker1975a, Ellis1985, TOUCHETTE20091}.

For LDT to be applied to practical problems requires the development of robust 
numerical tools to compute LDFs. 
Monte Carlo sampling methods, such as
the cloning algorithm and transition path sampling \cite{Prados:2016, 
nemoto2017finite, ray2018importance} augmented with importance sampling 
\cite{Nemoto2016,klymko2017rare,ray2018exact} as well as recent direct rate 
function evaluation techniques 
\cite{Whitelam2019} have been applied to lattice and continuum nonequilibrium 
systems~\cite{Derrida2003, Bodineau2005, garrahan2009first, Jan2011, 
gorissen2011finite,gorissen2012exact,Prados:2012,Prados:2016, Garrahan2018b, 
ray2018importance, Brewer2018, Schile2018, GrandPre2018, Gao2018}. 
Alternatively, tensor network (TN) methods 
provide analytic or numerical representations of the steady state of a master 
equation,
with some common examples being the matrix ansatz~\cite{derrida1993exact,blythe2007nonequilibrium} 
and the density matrix renormalization group (DMRG) algorithm
~\cite{carlon1999density,gorissen2009density,gorissen2012exact,
gorissen2011finite,PhysRevE.82.036702}.
Recently, the TN approach has been applied to kinetically constrained models 
of glasses~\cite{banuls2019using}.

In this report, we compute the current cumulant generating function and other 
properties of the dynamical phases of the
1D (single-lane) and multi-lane asymmetric simple exclusion process (ASEP), 
under open boundary conditions, using matrix product states (MPS) and the DMRG.
In 1D, the ASEP is a paradigmatic model of nonequilibrium statistical mechanics
that can be solved
semi-analytically with the matrix ansatz and functional Bethe ansatz and has 
been studied via many numerical 
approaches~\cite{derrida1998exact, prolhac2009combinatorial, 
lazarescu2015physicist, Lazarescu2014},
though detailing the microscopic structure in all regions of the phase diagram 
remains challenging.
While DMRG has previously been used to compute high-order current cumulants of 
the 1D ASEP to verify analytic expressions~\cite{gorissen2012exact}, 
a systematic application across the phase diagram of this model has yet to be 
presented. 
Thus we start with a short benchmarking study of the 1D ASEP using the DMRG, 
focusing
on the phases induced by a current bias and the associated macro and 
microscopic behaviors.
We then examine the phase behavior of the multi-lane version of the ASEP under 
a longitudinal current bias for systems with up to 4 lanes.
The multi-lane model can be thought of as a finite-width version of the 2D ASEP
and it is the first time, to our knowledge, that this family of models has been
studied.

\section{Large Deviation Theory and Matrix Product States}\label{methods}
We first briefly summarize some relevant concepts in large deviation theory, 
the theory of matrix product states
and the density matrix renormalization group.
A more complete description can be found in recent reviews~\cite{TOUCHETTE20091,
touchette2011basic,ray2018importance,schollwock2011dmrgmps}.

In a nonequilibrium system, the state vector $|P_t\rangle$ evolves from an 
initial state $|P_0\rangle$ according
to a master equation with dynamics generated by a non-Hermitian Markov operator
$\W$,
\begin{align}
	\partial_t |P_t\rangle= \W |P_t\rangle, 
\end{align}
with the probability of a system configuration $\mathcal{C}$ at time $t$ given 
by  $\text{Prob}(\mathcal{C}_t) \equiv \langle \mathcal{C}|P_t\rangle$.
The long-time limit yields the final (steady) state $|P_\infty\rangle$.
The probability of observing a given trajectory of configurations 
$\mathscr{C}(t_N)=\{\mathcal{C}_0,\mathcal{C}_1,\dots,\mathcal{C}_{t_N}\}$ at 
times $\{t_0,\cdots,t_{N}\}$ ($dt=t_N/N$) is,
\begin{align}
	\text{Prob}(\mathscr{C}(t_N)) = \text{Prob}(\mathcal{C}_0)
    \prod_{i=0}^{t_{N-1}} \langle \mathcal{C}_{i+1}| e^{dt\W} |
    \mathcal{C}_i\rangle.
\end{align}
We can define dynamical observables along such a trajectory, such as a 
time-local observable
${\mathcal{O}} = \sum_{i=0}^{t_{N-1}} o(\mathcal{C}_{i+1},\mathcal{C}_{i})$, 
with $o$ being an arbitrary function of time-adjacent configurations 
($\mathcal{C}_{i+1}$ and $\mathcal{C}_{i}$). 
To characterize the steady-state expectation value and fluctuations of this 
observable, we define a cumulant generating function, $\psi(\lambda)$,
\begin{align}
 \label{Eq:LDF}
\psi(\lambda) &= \lim_{t_N\rightarrow\infty} \ln \left <  e^{-\lambda 
    \mathcal{O}} \right > \nonumber\\
&=  \lim_{t_N\rightarrow\infty} \ln \sum_{\mathscr{C}(t_N)} \text{Prob}
    (\mathscr{C}(t_N)) e^{-\lambda \mathcal{O}}, 
\end{align}
where $\lambda$ is a field conjugate to the observable. 
At $\lambda=0$, the first derivative of $\psi$ is the observable's steady-state
expectation value $\langle o\rangle$; 
characterizations of the fluctuations of $o$, via its cumulants, are obtained 
from higher-order derivatives of $\psi$.
A fundamental result in LDT is that $\psi(\lambda)$
is the largest eigenvalue of a tilted operator $\W_\lambda$, i.e., 
 \begin{align}
 \W_\lambda |P^\lambda\rangle = \psi(\lambda)  |P^\lambda\rangle,
 \label{eq:maineq1}
 \end{align}
where, for discrete configurations, 
\ba
\W_\lambda(\mathcal{C},\mathcal{C}') = \W(\mathcal{C},\mathcal{C}')
e^{-\lambda o(\mathcal{C},\mathcal{C}')}(1-\delta_{\mathcal{C},\mathcal{C}'}) 
-R(\mathcal{C})\delta_{\mathcal{C},\mathcal{C}'}
\ea
with $R(\mathcal{C})=\sum_{\mathcal{C}\neq\mathcal{C}'} \W(\mathcal{C},
\mathcal{C}')$. 
Furthermore, the corresponding right (left) eigenvector $|P^\lambda\rangle$ 
($\langle P^\lambda|$) 
gives the probability of a configuration in the final (initial) state 
conditioned on trajectories satisfying 
$\langle o\rangle = d\psi(\lambda)/d\lambda$~\cite{Lebowitz1999}.

The computation of $\psi(\lambda)$ from the eigenvalue problem in 
Eq.~\eqref{eq:maineq1} can be recast
as a generalized variational problem,
  \begin{align}
 \langle \delta P^\lambda| \W_\lambda |P^\lambda \rangle - \psi(\lambda) 
      \langle \delta P^\lambda |P^\lambda\rangle = 0.
 \label{eq:gvp}
  \end{align}
Because $\W_\lambda$ is non-Hermitian, for an approximate $|P^\lambda\rangle$, 
$\langle P^\lambda|$, 
$ \langle P^\lambda| \W_\lambda |P^\lambda \rangle $ may be above or below the 
exact $\psi(\lambda)$.

In this work, we use an MPS as an ansatz
for $|P^\lambda\rangle$ and perform the optimization in Eq.~\eqref{eq:gvp} 
using the DMRG algorithm for
non-Hermitian operators~\cite{carlon1999density,chan2005density}. For a lattice
of length $L$,
a configuration $\mathcal{C}$ is an ordered list of the local states $\sigma_i$
of sites $i=1\ldots L$,
\begin{align}
|\mathcal{C}\rangle = \ket{\sigma_1,\cdots,\sigma_L}
\end{align}
An MPS expresses the configurational probability $\text{Prob}(\mathcal{C})$ as 
a matrix product
\ba
\text{Prob}(\mathcal{C}) =\matr{M}^{\sigma_1}\matr{M}^{\sigma_2}\cdots
\matr{M}^{\sigma_{L-1}}\matr{M}^{\sigma_L} \label{eq:mps}
\ea
where the matrices $\matr{M}^{\sigma_i}$ ($i = 2\ldots L-1$) are of dimension 
$D\times D$, and the first
and last matrices are of dimension $1\times D$ and $D\times 1$ respectively. 
The bond dimension $D$ controls the
accuracy of the ansatz and may be increased until the ansatz is exact. 
The matrix product contains a local gauge (i.e.
$\{ \matr{M}^{\sigma_i} \}$ can be varied while keeping the matrix product 
invariant) which can be fixed by choosing a canonical form,
\ba
\text{Prob}(\mathcal{C}) =\matr{L}^{\sigma_1}\matr{L}^{\sigma_2}\cdots
\matr{F}^{\sigma_i}\cdots\matr{R}^{\sigma_{L-1}}\matr{R}^{\sigma_L},
\label{eq:canon}
\ea
where $\sum_\sigma {\matr{L}^\sigma}^\dag \matr{L}^\sigma = \matr{I}$ and 
$\sum_\sigma \matr{R}^\sigma {\matr{R}^\sigma}^\dag = \matr{I}$.

The canonical form of Eq.~\eqref{eq:canon} also simplifies the computation of 
the bipartite entanglement entropy $S(i)$ at site $i$, 
a quantification of the non-factorizable correlations between the states of 
sites to the left and right of site $i$. 
By reshaping the central rank-3 tensor into a matrix, with 
$\matr{G}_{\sigma_ip,q}=\matr{F}^{\sigma_i}_{pq}$, $S(i)$ is conveniently 
computed as, 
\begin{align}
  S(i) = -\sum_m s_m^2 \log_2 s_m^2 \label{eq:eentropy},
  \end{align}
where \{$s_m$\} are the singular values of $\matr{G}$.

\begin{figure}
\centering
\includegraphics[width=0.8\linewidth]{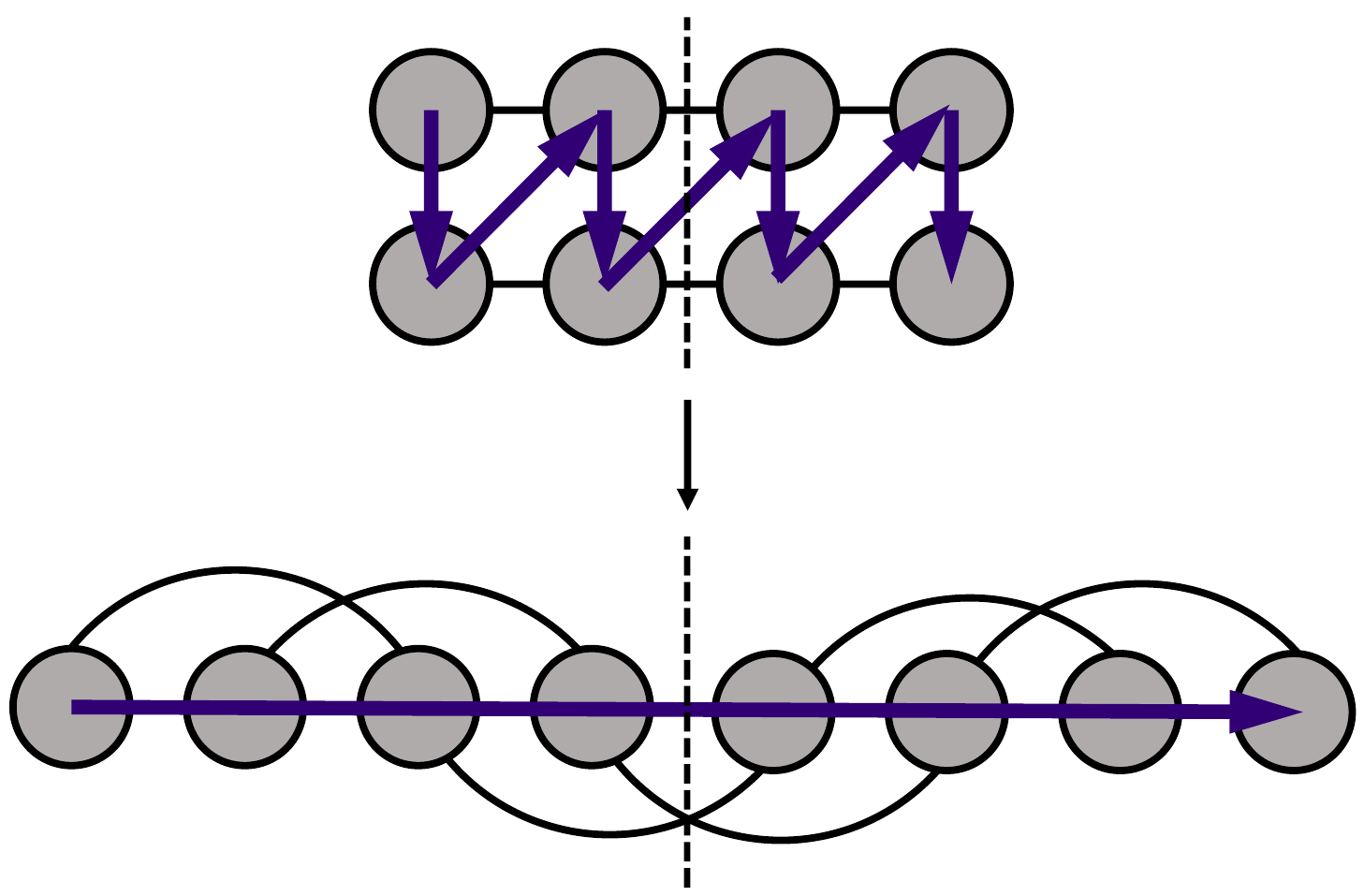}
\caption[MPO Snaking]{
A diagrammatic representation of the mapping of a 2D lattice with nearest 
    neighbor interactions onto a 1D lattice with long range interactions. 
The arrows indicate how our DMRG optimization traverses the 2D lattice
and the dashed line shows the bond over which the reported entanglement entropy
    is measured.
}
\label{fig:snake}
\end{figure}

Because the MPS representation of a state requires a 1D site ordering, 
associated with the sequence of matrices in Eq.~\eqref{eq:mps},
we must define a 1D traversal pattern for the multi-lane ASEP. We do so using
a zig-zag ordering of sites, shown in Fig.~\ref{fig:snake}. Note that for fixed
accuracy, the bond dimension of the matrices in the MPS 
usually needs to increase exponentially with the number of lanes studied, 
reflecting independent fluctuations in the different lanes.

\section{Model}\label{sec:model}

\begin{figure}
\centering
\includegraphics[width=0.99\linewidth]{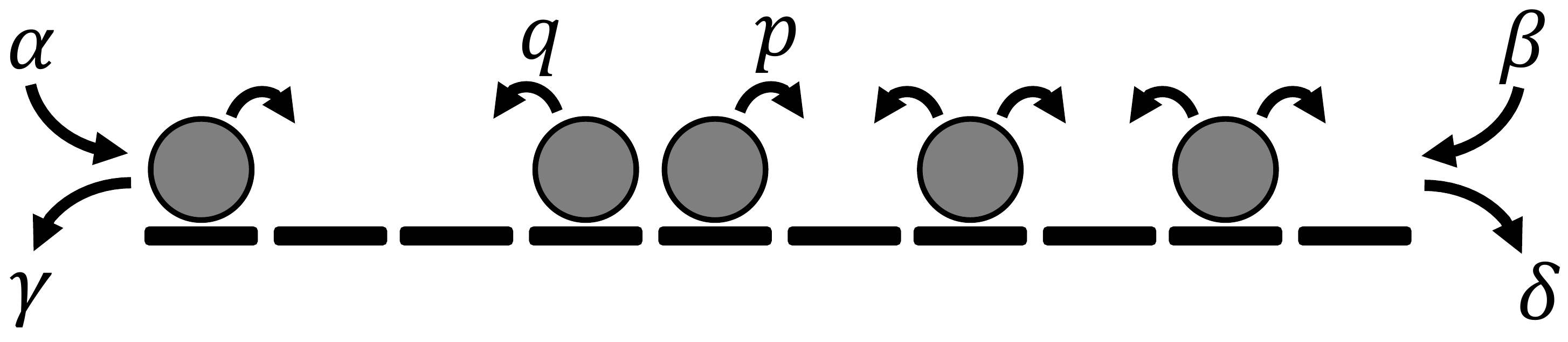}
\caption[ASEP Diagram]{The ASEP model where particles on a 1D lattice 
    stochastically hop to a vacant neighboring right (left) site 
at a rate of $p$ ($q$) and enter (exit) at the left and right boundaries at 
    rates $\alpha$ ($\gamma$) and $\beta$ ($\delta$).}
\label{fig:sep}
\end{figure}

The 1D Simple Exclusion Process (SEP) (Fig. \ref{fig:sep}) takes place on a 1D 
lattice of $L$ sites. 
Particles hop stochastically to vacant nearest-neighbor sites at the following 
rates.
In the lattice interior, particles hop right (left) with rate $p$ ($q$) with 
asymmetry enforced via $p\neq q$ (ASEP).
At the edges, particles enter (exit) at the left with rate $\alpha$ ($\gamma$) 
and at the right with rate $\beta$ ($\delta$).
In this work we focus on phases induced
by the current bias $\lambda$ in the parameter regime 
$\alpha=\beta=\gamma=\delta=1/2$ and $p + q = 1$.
The tilted operator for the current cumulant generating function is,
\ba
\W^\text{1D}_\lambda =& \alpha\left(e^{\lambda}\matr{a}_1^\dag-\matr{v}_1\right)
+\gamma\left(e^{-\lambda}\matr{a}_1-\matr{n}_1\right)\\
&+\sum_{i=1}^{L-1}p\left(e^{\lambda}\matr{a}_i\matr{a}_{i+1}^\dag-\matr{n}_i
\matr{v}_{i+1}\right)\\
&+\sum_{i=1}^{L-1}q\left(e^{-\lambda}\matr{a}_i^\dag\matr{a}_{i+1}-\matr{v}_i
\matr{n}_{i+1}\right)\\
&+\beta\left(e^{-\lambda}\matr{a}_L^\dag-\matr{v}_L\right)+
\delta\left(e^{\lambda}\matr{a}_L-\matr{n}_L\right),
\ea
where $\matr{a}_i$, $\matr{a}_i^\dag$, $\matr{n}_i$ and $\matr{v}_i$ are 
annihilation, creation, particle number, and vacancy number operators.
Note that the tilted operator is invariant with respect to the combined 
operation of particle-hole transformation/inversion 
($\matr{a}^\dagger\leftrightarrow\matr{a}$ and $\{...,i,i+1,...\}
\leftrightarrow\{...,i+1,i,...\}$). 
The eigenvalues of $\W^\text{1D}_\lambda$ also exhibit a Gallavotti-Cohen (GC) 
symmetry~\cite{gallavotti1995dynamical,Lebowitz1999} of the form 
$\psi(\lambda)=\psi(\lambda^*)$ where,  for the specified ASEP parameters, 
$\lambda^* = -\frac{L-1}{L+1}\ln(p/q)-\lambda$. 

The multi-lane ASEP is defined on a 2D lattice of $L_y \times L_x$ sites. 
It augments the 1D ASEP with bulk hopping in the vertical (transverse) 
direction (at rates $p_y$, $q_y$) and particles inserted and removed at the 
vertical boundaries (at rates $\alpha_y, \beta_y, \gamma_y, \delta_y$).
We apply the current bias in the (longitudinal) $x$-direction, with a tilted 
operator that takes the form,
\begin{align}
  \W^\mathrm{2D}_\lambda = \W^\mathrm{1D_x}_\lambda + \W^\mathrm{1D_y}_0,
\end{align}
and retains the above GC and particle-hole/inversion symmetries.
To understand the effects of the transverse parameters on the longitudinal 
system's phase behavior,
we focus on two multi-lane parameter sets, namely open and closed vertical 
boundaries. 
Both require $p_x+q_x=1$, $p_y=q_y=1/2$, and 
$\alpha_x=\beta_x=\gamma_x=\delta_x=1/2$, 
while the open (closed) case specifies $\alpha_y=\beta_y=\gamma_y=\delta_y=1/2$
($\alpha_y=\beta_y=\gamma_y=\delta_y=0$).

To characterize the system, the DMRG algorithm is used to determine the largest
eigenvalue of each tilted operator, through which 
the steady-state total current and current susceptibility are computed as 
$J=\partial \psi(\lambda)/d\lambda$ and 
$\chi=\partial^2 \psi(\lambda)/d\lambda^2$.
Local densities, currents, and activities may also be computed by contracting 
the resulting left and right eigenvector with the appropriate operator, i.e.,
\ba\label{eq:obs}
\rho_i&=\bra{P^\lambda}\matr{n}_i\ket{P^\lambda},\\
J_i&=\bra{P^\lambda}pe^\lambda\matr{a}_i\matr{a}_{i+1}^\dag-qe^{-\lambda}
\matr{a}_i^\dag\matr{a}_{i+1}\ket{P^\lambda},\\
K_i&=\bra{P^\lambda}pe^\lambda\matr{a}_i\matr{a}_{i+1}^\dag+qe^{-\lambda}
\matr{a}_i^\dag\matr{a}_{i+1}\ket{P^\lambda},
\ea
assuming $\bra{P^\lambda}\ket{P^\lambda}=1$.

\section{Results}\label{results}

\begin{figure}
\centering
\includegraphics{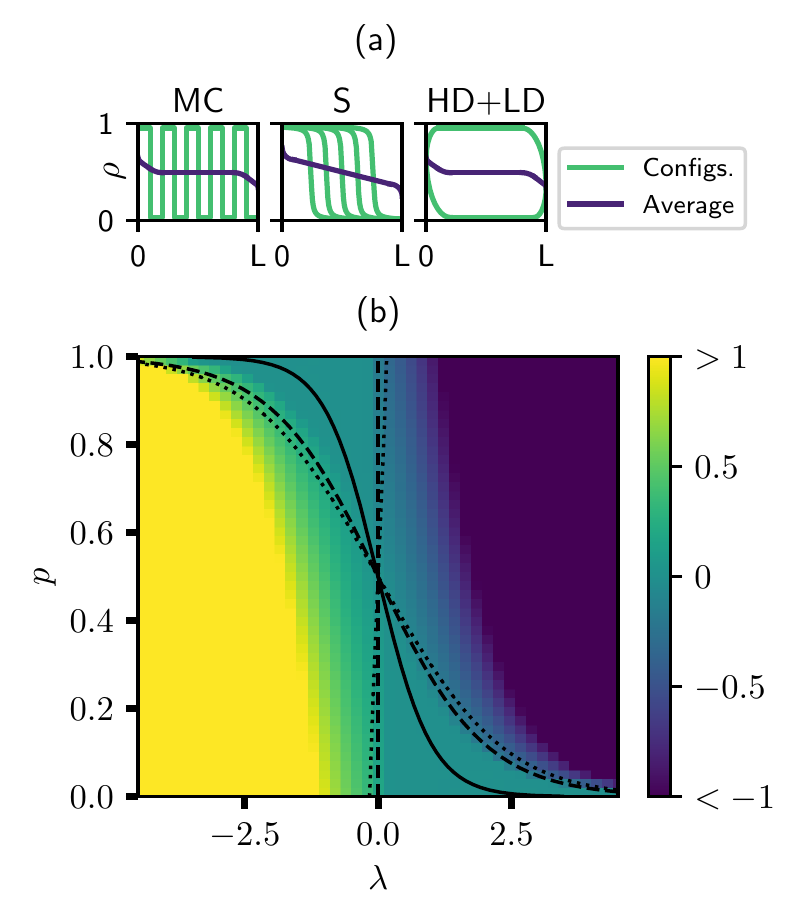}
\caption[Dynamical Phase Diagram]{
(a) Rudimentary sketches of the density profiles in the three possible phases. 
Blue curves represent approximate steady-state density profiles while green 
    curves depict typical particle configurations.
(b) A map of the dynamical phase behavior of the ASEP showing the steady-state 
    current $J$ as a function of $p$ and $\lambda$ for a length $L=20$ lattice 
    as determined via DMRG.
Additionally shown in black are lines indicating the center of the GC symmetry 
    (solid) and the predicted boundaries between the MC and shock phases 
    (dotted, via macroscopic fluctuation theory~\cite{Bertini2002,
    bertini2015macroscopic,lazarescu2015physicist}) and the shock and HD+LD 
    phases (dashed, via functional Bethe ansatz~\cite{prolhac2008current,
    lazarescu2015physicist}).}
\label{fig:phases}
\end{figure}

\subsection{Benchmark MPS calculations of the 1D ASEP}
\begin{figure*}[t]
\centering
\includegraphics{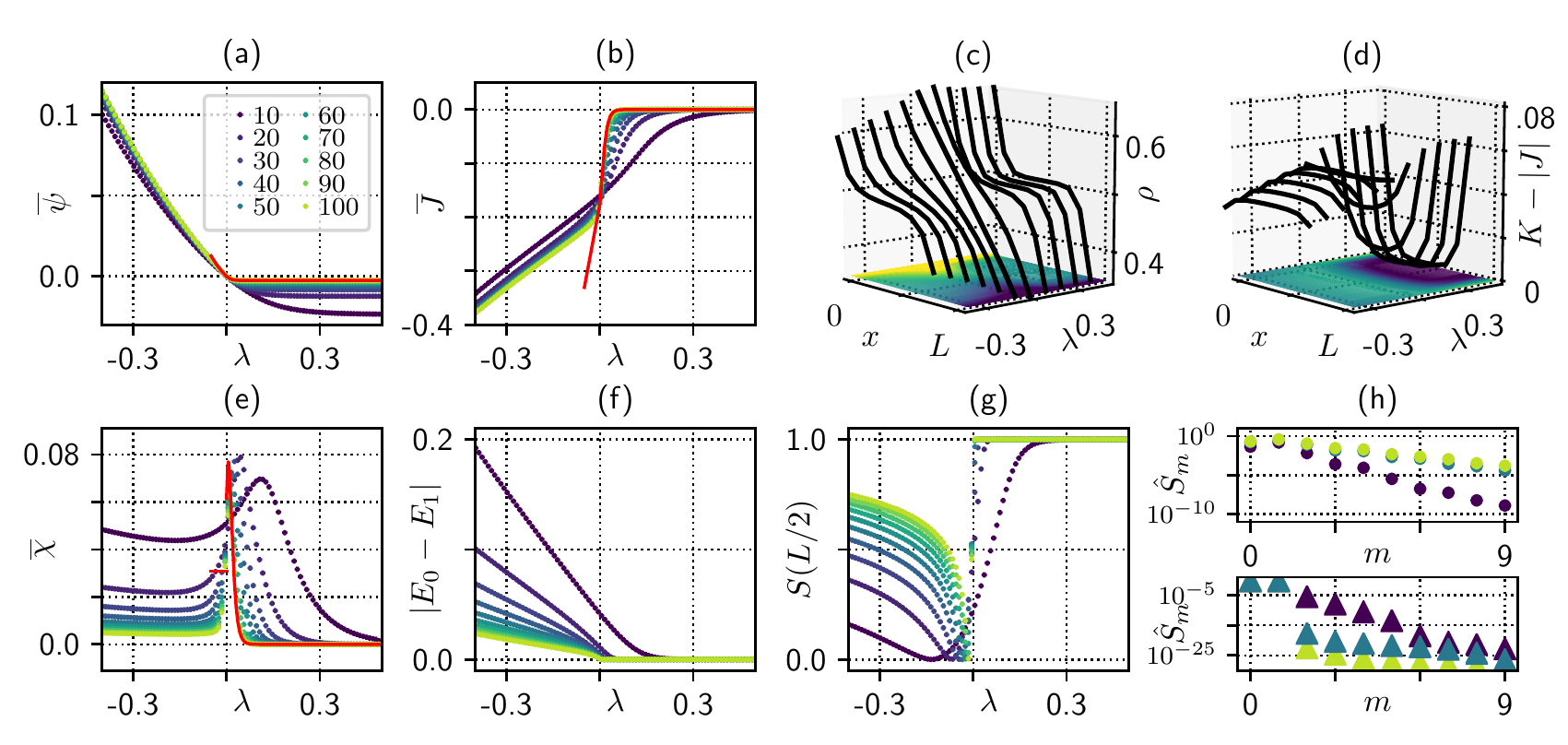}
\caption[1D DMRG Benchmark]{The behavior of the 1D ASEP with lattice lengths of
    $L=[10,100]$. The DMRG results for the normalized (a) CGF 
    $\overline{\psi}=E_0/L$, (b), current $\overline{J}=\partial_\lambda\psi/L$,
    and (e) current susceptibility $\overline{\chi}=\partial_\lambda\psi/L^2$ 
    compared with the analytic functional Bethe ansatz expressions (red), valid
    for $\lambda\to0^-$ and $\lambda>0$; 
additionally (f) shows the gap between the ground and first excited state 
    energies. 
Plots (c) and (d) show the density $\rho$ and recurrent hopping $K-|J|$ as a 
    function of position in a $L=10$ lattice, $x$, and $\lambda$. 
(g) shows the entanglement entropy $S$ of a bipartition at the center bond as 
    a function of $\lambda$ with the upper (lower) subfigures in (h) showing 
    the corresponding ordered entanglement spectrum, with 
    $\hat{S}_m=-s_m^2\log_2s_m^2$, at $\lambda = -0.3$ ($\lambda=0.3$).}
\label{fig:1D_size}
\end{figure*}

We begin by using MPS and DMRG to characterize the phase behavior in the 
aforementioned parameter space and benchmark this approach
against earlier results from the semi-analytical functional Bethe ansatz and 
the approximate macroscopic fluctuation theory~\cite{lazarescu2015physicist}.
In this space, there are three expected phases, which are described in 
Fig.~\ref{fig:phases}(a) via rudimentary sketches of both the steady-state 
density profile and the most probable particle configurations.
These are the Maximal Current (MC) phase, where, in the most probable 
microscopic configurations, particles are evenly spaced throughout the lattice,
allowing a maximal amount of biased hopping, 
the Shock (S) phase, where particles conglomerate on one side of the lattice 
to form a shock that, in path-space simulations, performs a Brownian walk on 
the lattice, 
and the High-Density/Low-Density Coexistence (HD+LD) phase, where the entirely 
filled and empty states (with some boundary effects) are degenerate in the 
thermodynamic limit and correspond to a steady-state density profile of 
$\rho=1/2$.

The predicted phase diagram is mapped in Fig.~\ref{fig:phases}(b) where the 
lines indicate the line of GC symmetry (solid), the boundary between the MC and
S phases (dotted, via macroscopic fluctuation theory), and the boundary between
the S and HD+LD phases (dashed, via functional Bethe ansatz). 
The steady-state current is also shown, computed via DMRG for an $L=20$ ASEP, 
showing that current functions as a dynamical order parameter for the 
transition from S to HD+LD, going effectively to zero in the HD+LD phase.
While the boundary between the MC and S phases is commonly defined as the point
where the per site current is $J=1/4$, we are not aware of an order parameter 
for this transition, which instead appears as a smooth crossover in the current
rather than a true phase
boundary.
Also note that because of the symmetries of the system, the remaining analysis 
can be limited to the lower left region of the parameter space ($p<1/2$ to the 
left of the line of GC symmetry), with the rest of the diagram mapped out by 
symmetry. 

Finite size errors can be converged rapidly by increasing the lattice size.
In Fig.~\ref{fig:1D_size}, we characterize this behavior using system 
properties such as the cumulant generating function, current, current 
susceptibility, and excited state gap for a range of $\lambda$ near $\lambda=0$
with $p=0.1$ and for lattice sizes up to $L=100$ via DMRG with bond dimension 
$D$ between $50$ and $300$. 
As a benchmark, the solid red line in Fig.~\ref{fig:1D_size} (a), (b), and (e) 
corresponds to the functional Bethe ansatz results, 
which is valid only in the HD+LD phase and near $\lambda=0$ in the S phase.

As $L\to\infty$, a number of interesting behaviors are observed, particularly 
at the interface between the S and HD+LD phases.
In this region, the cumulant generating function transitions from having a 
finite negative slope to become nearly flat, signifying a transition to a low 
current regime.
We also see that the system becomes gapless here due to the degeneracy of the 
high-density and low-density configurations. 
Because the two degenerate states are of the same particle-hole/inversion 
symmetry while $\partial_\lambda\W_\lambda$ is odd under this symmetry, the 
closing gap does not contribute to the spike in the current susceptibility.

The MPS representation also provides the state's full configurational 
information, enabling us to study the microscopic structure of the phases
and quantities that are not derivatives of the cumulant generating function.
Fig.~\ref{fig:1D_size} (c) and (d) show the steady-state density, $\rho$, and 
recurrent hopping, $K-|J|$, computed as specified in Eq.~\eqref{eq:obs}, as a 
function of the position in the lattice $x$ and the current bias $\lambda$. 
These density profiles correspond to those shown in Fig.~\ref{fig:phases}(a), 
with the linear profile near $\lambda=0$ corresponding to the shock phase.
The HD+LD and MC phases can here be distinguished via the rate of recurrent 
hopping; 
particles and holes are spatially dispersed in the MC phase, allowing frequent 
opportunities to hop back and forth, as indicated by the finite observed 
recurrent hopping at $\lambda<0$. 
When the transition is made into the HD+LD phase, the recurrent hopping drops 
to nearly zero in the lattice bulk, attributable to the lattice being nearly 
entirely filled or empty in this phase and thus providing few opportunities 
for recurrent hops. 

An additional way to summarize the microscopic information (and the associated 
correlations in the system) is via the 
entanglement entropy and entanglement spectrum ($S(i)$ and $\{ s_m \}$ in 
Eq.~\eqref{eq:eentropy}) which we measure at the middle of the lattice. 
These are plotted for the right eigenvector $|P^\lambda\rangle$ in 
Fig.~\ref{fig:1D_size} (g).
The entanglement spectrum provides details on the maximum bond dimension 
required to accurately represent a state and can be used as a generalized 
order parameter~\cite{kitaev2006topological,pollmann2010entanglement}. 
There are two clear regions present in the entanglement entropy, one 
corresponding to the MC phase, the other to the HD+LD phase.
For the MC phase, the spectrum decays slowly, indicating that a relatively 
large bond dimension is required to accurately represent the given state.
In the HD+LD phase, the entanglement entropy is larger and appears to be 
exactly $1$ ($\log_2 2$). 
The entanglement spectrum shows that only two modes contribute, arising
  from the filled and empty configurations, indicating the state can be 
  represented exactly by an MPS of bond dimension $2$.
  It is evident that the entanglement entropy converges as a function of $L$, 
  indicating an area law. 

\subsection{Multi-lane ASEP model}

We now consider a system comprised of multiple ASEP lanes, with particles that 
may hop vertically ($y$-direction) or horizontally ($x$-direction),
where we will examine the unexplored interplay between vertical and horizontal 
currents that can generate new phase behaviour.

\subsubsection{Closed Multi-lane ASEP}

\begin{figure*}[t]
\centering
\includegraphics{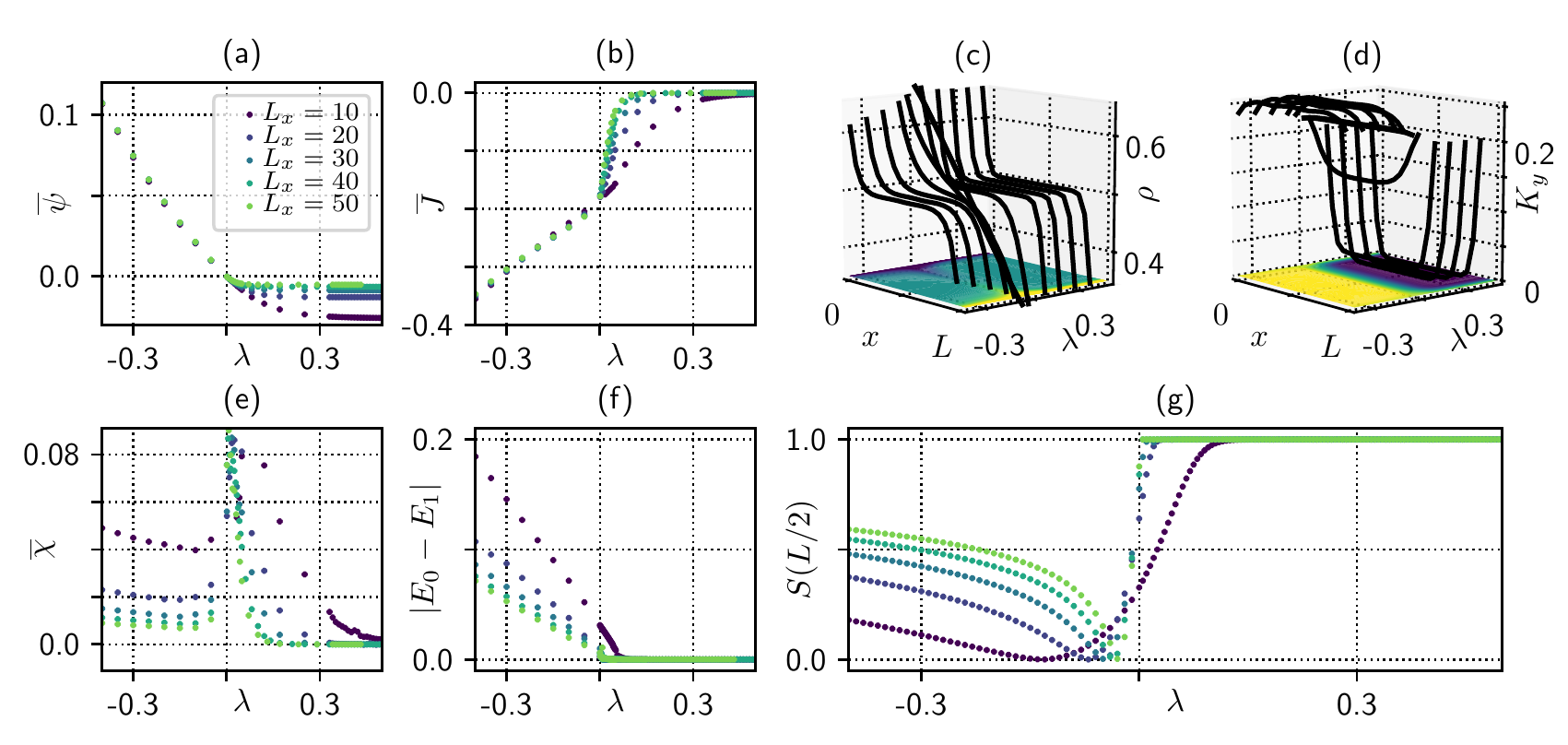}
\caption[Closed Multi-lane ASEP Results]{The behavior of the closed multi-lane 
    ASEP showing the DMRG results for the normalized (a) CGF 
    $\overline{\psi}=E_0/(L_xL_y)$, (b) current 
    $\overline{J}=\partial_\lambda\psi/(L_xL_y)$, and (e) current 
    susceptibility $\overline{\chi}=\partial_\lambda^2\psi/(L_x^2L_y)$
as well as (f) the gap between the ground and first excited state energies for 
    the four lane systems with lengths up to $L_x=50$. 
Plots (c) and (d) show the density $\rho$ and vertical hopping activity $K_y$ 
    between lanes for a two-lane ASEP with $L_x=20$. 
(g) Shows the entanglement entropy $S$ of a bipartition of the system at the 
    center bond as a function of $\lambda$. 
}
\label{fig:closed}
\end{figure*}

\begin{figure*}[t]
\centering
\includegraphics{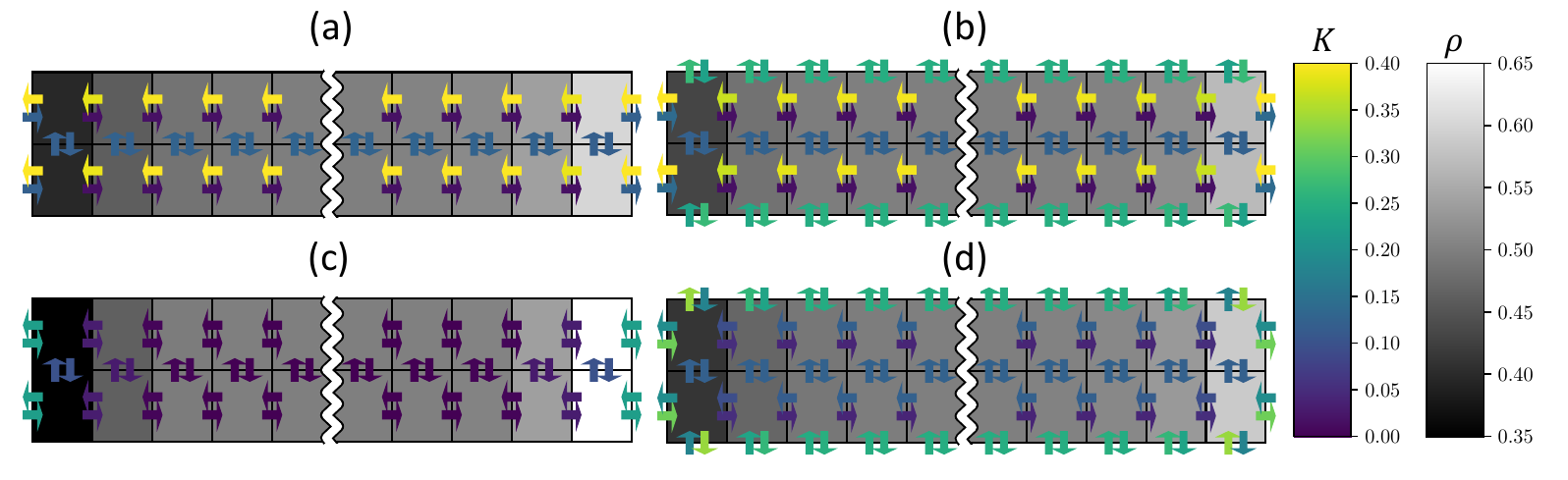}
\caption[Comparison of two-lane microscopic configurations]{
Representations of the measured microscopic observables for a lattice of size 
    $2\times20$ for systems with 
both closed vertical boundary conditions, in (a) the MC phase ($\lambda=-0.3$) 
    and (c) the HD+LD phase ($\lambda=0.3$), 
and open vertical boundary conditions, in (b) the MC phase ($\lambda=-0.3$) 
    and (d) the HD+LD phase ($\lambda=0.3$). 
The grayscale shading of lattice sites indicates the measured steady-state 
    density $\rho$ while the color of the arrows corresponds to the 
    steady-state number of hops $K$ in the given direction.
Only the five edge sites on each side are included because the behavior in the 
    lattice bulk resembles what is seen at the center-most included sites. 
}
\label{fig:configs}
\end{figure*}

A simple, but nontrivial, extension of the 1D ASEP into multiple lanes, as 
specified in Sec.~\ref{sec:model}, is to augment horizontal hopping and 
entry/exit parameters
with equal vertical hopping rates $p_y=q_y=1/2$ and no entry/exit at the 
vertical bounds, i.e. closed boundary conditions. 
To understand the phase behavior here, we again carried out DMRG calculations 
mapping out the behavior as a function of the longitudinal current bias 
$\lambda_x$ for fixed $p_x=0.1$, with bond dimensions $D$ between $50$ and 
$300$ and with system widths and lengths of up to $L_y=4$ and $L_x=50$. 

The resulting cumulant generating function, current, current susceptibility, 
and first excited state gap are displayed respectively in Fig.~\ref{fig:closed}
(a), (b), (d) and (e) for the $L_y=4$ ASEP (with the $L_y=[2,3]$ results being 
essentially indistinguishable from these).
A comparison between this figure and Fig.~\ref{fig:1D_size} shows no 
qualitative difference between the single lane and closed multi-lane ASEP. 
We can analyze the ground state MPS to confirm whether the microscopic 
configurations in the multi-lane system correspond to those seen in 1D.

Fig.~\ref{fig:closed} (c) and (d) show the behaviors of key observables as a 
function of $\lambda$ while Fig.~\ref{fig:configs} (a) and (c) compare 
snapshots of microscopic observable behaviors when the system is respectively 
in the MC and HD+LD phases. 
Using results from a two lane calculation, Fig.~\ref{fig:closed} (c) shows the 
density profile in one of the lanes as a function of $\lambda$, with the most 
notable point being the linear profile near $\lambda=0$, indicative of a shock 
phase. 
The MC and HD+LD phases are again indistinguishable by their density profiles, 
emphasized in Fig.~\ref{fig:configs} (a) and (c) where the steady-state density
corresponds to the shading of the lattice sites. 
As an means of distinguishing the two phases, we can use either the horizontal 
recurrent hopping rate profile (as done in 1D and not shown here) or the 
vertical activities between the two lanes as demonstrated in both 
Fig.~\ref{fig:closed} (d) and a comparison of Fig.~\ref{fig:configs} (a) and 
(c). 
Here, the bulk vertical activity is near $K_y=1/4$ per site when in the MC 
phase, supporting a microscopic structure where particles neighbor holes with 
probability $1/2$ and the probability of a vertical hop when such a 
configuration occurs is $p_y=q_y=1/2$. 
After crossing the 1D ASEP phase boundary at $\lambda=0$, the bulk vertical 
activity approaches zero, indicating that hops are prevented by an entirely 
full or empty lattice and demonstrated by the lack of any hopping shown in 
Fig.~\ref{fig:configs} (c). 

This picture is further supported by the profile of the entanglement entropy 
for the two-lane ASEP shown in Fig.~\ref{fig:closed} (g),
which again mimics the behavior seen for the 1D ASEP. 
If an area law holds, the entanglement entropy across the central cut should 
grow linearly with the width of the system.
Instead, here the similarity to the 1D profile arises because the
HD+LD phase results from entirely empty and full configurations (where
particle occupancy is perfectly correlated between the two lanes in both
configurations).
Also, similarly to in 1D, the entanglement entropy converges as a function of 
lattice length $L_x$.

\subsubsection{Open Multi-lane ASEP}
\begin{figure*}[t]
\centering
\includegraphics{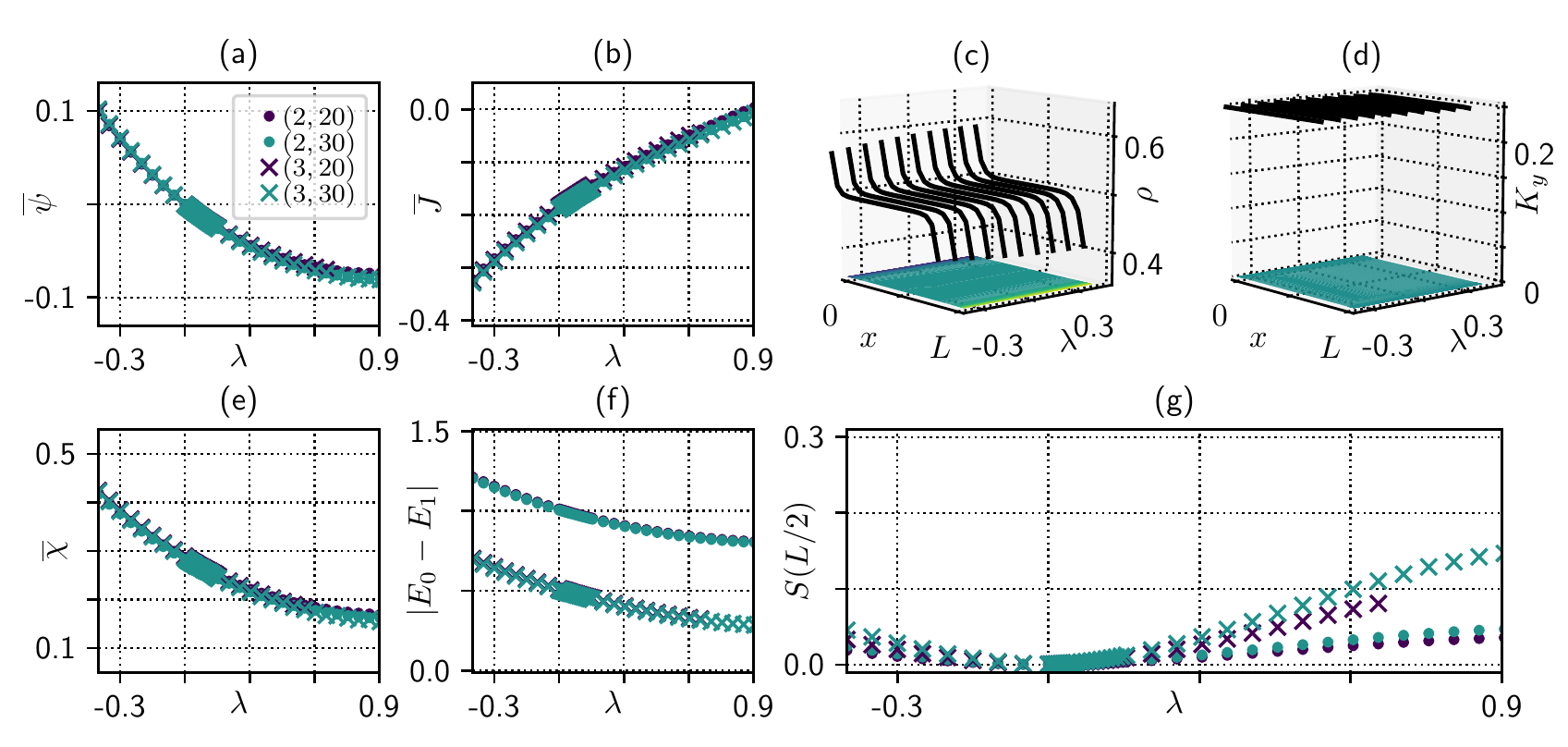}
\caption[Open Multi-lane ASEP Results]{The behavior of the open multi-lane 
    ASEP showing the DMRG results for the normalized (a) CGF 
    $\overline{\psi}=E_0/(L_xL_y)$, (b) current 
    $\overline{J}=\partial_\lambda\psi/(L_xL_y)$, and (e) current 
    susceptibility $\overline{\chi}=\partial_\lambda^2\psi/(L_xL_y)$
as well as (f) the gap between the ground and first excited state energies for 
    the two- and three-lane systems with lengths up to $L_x=30$. 
Plots (c) and (d) show the density $\rho$ and vertical hopping activity $K_y$ 
    between lanes for a two-lane ASEP with $L_x=20$. 
(g) shows the entanglement entropy $S$ of a bipartition of the system at the 
    center bond as a function of $\lambda$. 
}
\label{fig:open}
\end{figure*}

To quantify the effects of vertical boundaries on the horizontally biased 
dynamical phase behavior of this multi-lane ASEP, we further consider a 
vertically open multi-lane ASEP,  
where vertical entry/exit rates are $1/2$, as specified in Sec.~\ref{sec:model}.
In these calculations, we employed DMRG to study the ASEP behavior as a 
function of the horizontal bias, $\lambda_x$, near $\lambda_x=0$, with 
$p_x=0.1$ for systems of up to length $L_x=50$ with up to three lanes ($L_y=3$)
using a maximum bond dimension of $D=50$. 

The results are displayed in Fig.~\ref{fig:open}, with the cumulant generating 
function, current, current susceptibility, and first excited state gap being 
shown in subfigures (a), (b), (e), and (f). 
The per site macroscopic observables are nearly indistinguishable for the 
various system sizes, 
with the only noticeable difference caused by the requisite shifting of the 
point of GC symmetry as a function of system length. 
While in the closed multi-lane model the current detected a transition into 
the HD+LD phase, no such transition is apparent here.

This is further supported by a microscopic analysis for a lattice of size 
$2\times20$.
The density and activity profiles are shown in Fig.~\ref{fig:open} (c) and (d) 
as a function of $\lambda$ and with more detailed observable information in the
snapshots at $\lambda=-0.3$ and $\lambda=0.3$, shown respectively in 
Fig.~\ref{fig:configs} (b) and (d).
The $\lambda$ sweep show no changes in the behavior of the density and vertical
activity. This is also true at $\lambda=0$, where the phase transition would be
expected to occur. 
The snapshots show no distinguishable differences in the density profiles in 
the regions where the MC and HD+LD phases would be expected. 
While the steady-state number of hops between lattice sites does not seem to 
indicate any phase transition, evidenced in the snapshots by comparisons of the
vertical hopping rates between lanes, we note that the desired low current 
behavior is created in a MC-like density profile by causing a small current to 
flow to the left in the bulk to counter the large current flowing to the right 
at the boundaries. 
This also illustrates a significant difference between the single-lane and 
multi-lane systems, namely that the steady-state current need not be spatially 
homogenous. 

The lack of the phase transition in the open multi-lane system contrasts with 
the behaviour of the closed multi-lane system.
The behavior of the open model likely arises due to the availability of a vertical 
particle bath that enables rapid relaxation when jammed phases begin to form.

\section{Conclusions}\label{conc}

In conclusion, we have used MPS and DMRG to conduct a systematic study of the 
1D and multi-lane ASEP with open horizontal
boundary conditions under a current bias.
In addition to providing a simple numerical route to compute macroscopic 
quantities out of equilibrium,
such as the cumulant generating function and its derivatives, these methods 
also provide access
to details of the underlying microscopic configurations. 
We find that the entanglement entropy and spectrum provides a global summary
of the correlations in the system, identifying the sharp structure of the 
transition into the HD+LD phase in the 1D ASEP. 
This transition is additionally marked by changes in the steady-state density 
and activity profiles.
In the case of the multi-lane ASEP, 
we find that the shock and HD+LD phases develop when vertical particle 
entry/exit is prohibited,
 but the phase boundary disappears entirely when this is reintroduced. 
This emphasizes the complex interplay between vertical and horizontal hopping 
parameters in this class of boundary driven processes. 

The MPS and DMRG are numerical realizations of the matrix ansatz
method that has long been used to produce semi-analytical solutions in driven 
lattice models. 
As this and other recent work shows~\cite{banuls2019using}, 
the flexibility of the purely numerical approach allows this framework to be 
applied to problems where
analytical techniques are difficult to use, such as the multi-lane ASEP. In 
addition, more general tensor network approaches
beyond MPS and DMRG allow for a natural treatment of two-dimensional, 
three-dimensional, and thermodynamic lattice systems~\cite{verstraete2008matrix,
orus2014practical,phien2015infinite}.
Applying these to two- and three-dimensional nonequilibrium statistical models
is an exciting possibility in the future.

\begin{acknowledgments}
This work was supported primarily by the US National Science Foundation (NSF) 
    via grant CHE-1665333. 
PH was also supported by a NSF Graduate Research Fellowship under grant 
    DGE-1745301 and an ARCS Foundation Award. 
\end{acknowledgments}
\bibliography{main}
\end{document}